# Statistical electricity price forecasting: A structural approach

*Raffaele Sgarlato*

**Abstract**

The availability of historical data related to electricity day-ahead prices and to the underlying price formation process is limited. In addition, the electricity market in Europe is facing a rapid transformation, which limits the representativeness of older observations for predictive purposes. On the other hand, machine learning methods that gained traction also in the domain of electricity price forecasting typically require large amounts of data. This study analyses the effectiveness of encoding well-established domain knowledge to mitigate the need for large training datasets. The domain knowledge is incorporated by imposing a structure on the price forecasting problem; the resulting accuracy gains are quantified in an experiment. Compared to an "unstructured" purely statistical model, it is shown that introducing intermediate quantity forecasts of load, renewable infeed, and cross-border exchange, paired with the estimation of supply curves, can result in a NRMSE reduction by 0.1 during daytime hours. The statistically most significant improvements are achieved in the first day of the forecasting horizon when a purely statistical model is combined with structured models. Finally, results are evaluated and interpreted with regard to the dynamic market conditions observed in Europe during the experiment period (1$^{st}$ October 2022 – 30$^{th}$ April 2023), highlighting the adaptive nature of models that are trained on shorter timescales.

**Table of contents**

# 1  Introduction

**Limited data availability**

In the European day-ahead market, electricity is traded in daily auctions with hourly granularity. The availability of historical prices as well as market related data is limited. For example, data provided by ENTSO-E Transparency only reach back to 2015. If prices are interpreted as a continuous hourly timeseries, thus disregarding the market design consisting of daily auctions, models can be trained on hourly prices rendering 8760 observations per year. However, power prices are characterized by strong daily patterns, and multivariate formulations are shown to be more accurate in practice [1]. In a multivariate setting, the number of observations per year drops to 365 daily auctions. Not only is the availability of historical data a limiting factor, but even more crucially, market conditions might vary substantially over time. The longer the period spanned by the training data, the more likely it is for the training data to include structural market changes and, ultimately, for the model to be trained on outdated market conditions. For example, a structural change (a reduction of available generation capacity) can hinder a model from correctly capturing the relationship between available capacity and electricity prices [2]. Longer training samples do not necessarily lead to more accurate forecasts since they can make the model slow in adapting to changing market conditions [3]. Such adaptation is possible, provided computational complexity allows for frequent retraining in the first place, which in the context of simpler models is shown to be desirable [4].

**Challenge for emerging methods**

Various machine learning applications emerged in the energy sector especially in recent years, which is reflected by the increasing number of publications on electricity demand, renewable production and price forecasting [5]–[7]. However, machine learning methods typically require an exceptional number of training observations [8], [9]. This limitation becomes apparent in studies that propose price forecasting models based on machine learning approaches but fall back on a rather small feature space. For example, data spanning over a decade has been used to model the Spanish day-ahead price with a univariate tree-based approaches with only 22 features [10]. One full year of hourly data has been used to train a convolutional neural network and forecast day-ahead prices only based on the input from lagged prices [11].[1] When larger feature spaces are used, more data are needed to train these models. It is shown that significant forecasting improvements can be achieved when information from neighbouring markets is incorporated, or even when forecasting prices of adjacent markets are produced simultaneously [2]. However, this finding in particular required 7 years of data and a complex feature selection procedure. Furthermore, the available capacity had to be dropped as an explanatory variable, since the relationship captured by the model did not generalize beyond the training set due to changing market conditions. A comprehensive review of algorithms [9] shows that machine learning models, and deep neural networks (DNN) in particular, can outperform traditional statistical approaches whilst incorporating up to 750 features – provided 7 years of data are available.[2]

To tackle the limited data availability, transfer learning and market integration have been proposed, and benefits have been observed under extremely simplified[3] conditions [12]. Whether the observed

---

[1] The study explores the predictive power of prices from all PJM zone for forecast the day-ahead prices of one PJM zone in particular.
[2] Of the 7 years, 5 years are used for the model training, 1 year for hyperparameter calibration, and 1 year for testing.
[3] Inputs of the reference study consisted only of lagged temperature, lagged prices, and weekday dummies

benefits of transfer learning materialize when given a richer and more realistic set of inputs remains an open question. Furthermore, this approach builds on the idea that these markets exhibit common "learnable" patterns. It is worth highlighting that "similar" markets are likely to contain redundant information, whereas "different" ones require additional features to characterize the difference.[4]

Longer training samples, transfer learning as well as univariate approaches require more relationships to be endogenous to the model (respectively explaining the difference between periods, between markets, or between hours of the day). The more relationships are required to be endogenous, the higher the model complexity and, therefore, the need for even more training data. The complexity that translates into more data being required – potentially hitting the boundaries imposed by the data availability – highlights the trade-off between the model's theoretical ability to recognise patterns and the deriving estimation uncertainty.

**Using domain knowledge**

A model's requirements on training data depend on the modelling approach. So-called fundamental or structural electricity system models[5] capture the basic physical and economic relationships with their structure and (strictly speaking) require no training data at all. This is not to say that historical data do not play a role. Historical data are still used to benchmark such models; however, they are not an integral part of the model. This is achieved by encoding domain knowledge into the model formulation. For instance, electricity system models typically include a representation of the demand and supply curves. These two curves are then used to compute the hourly marginal system cost which, under simplified assumptions, can be interpreted as the day-ahead's pay-as-clear price (marginal pricing).[6,7] Structural electricity system models are widely used to forecast prices, especially when the forecasting horizon reaches years, or even several decades into the future. The desired forecast horizon is arguably one of the main factors determining whether a purely statistical or a purely structural approach is used for electricity price forecasting. In the context of medium term price forecasting, a hybrid approach has been proposed consisting of a structural model complemented by a statistical procedure [14]. This mirrors the scope of this study that proposes a statistical model complemented by a structural procedure.

Encoding domain knowledge into a modelling approach is not peculiar to structural electricity system modelling. Domain knowledge is already used when choosing model inputs. Some studies [15] highlight the "fundamental" relationship between selected inputs, such as forecasted demand and capacity margin indicators, and the price formation process. Representing the price formation process in a "fundamental" fashion by estimating a price-demand function[8] (that then translates the demand forecast into a price forecast) is shown to be a promising approach [16]. Several advantages are mentioned, including: (i) the formulation allows to bypass the information loss caused by multivariate formulations,[9] (ii) statistical properties, such as seasonalities, can be attributed to the underlying demand rather than the estimated price-demand function, and (iii) individual component such as the input demand can be delegated to specialized models.

---

[4] For example, geographically close bidding zones are characterized by the similar (provided there is free NTC) and similar weather conditions (redundant information). On the other hand, the price formation in regions characterised by a high share of hydroelectric generation is fundamentally different to one characterized by a high penetration of variable renewable sources (different price formation patterns).
[5] For further details on the model categorization, consult Weron [13]
[6] Marginal system cost with respect to the energy balance equation.
[7] Simplified assumptions typically include, for example, perfect competition, linearity, perfect foresight.
[8] The proposed price-demand function is the counterpart to the merit order model.
[9] The 24-hour structure of the daily auctions is generally either adapted to a univariate approach, or represented by a multivariate approach, which is found to work better in practice [1].

**Structuring a statistical model**

Domain knowledge can be further leveraged by statistical methods by choosing an appropriate functional form, by an informed pre-selection of input parameters, or by constraining parameter coefficients estimate to "sensible" intervals. For example, when electricity prices are forecasted using wind speed directly, the factors that let wind speed contribute to the price formation are endogenous to the model. Although this theoretically allows to capture all sorts of relationships,[10] it often requires an exceptionally large feature space [17]. A large feature space increases the risk of overfitting and harms the accuracy of the model when accidental correlations in the training set overshadow causal relationships. Furthermore, a large feature space does not only necessitate a more stringent regularization to avoid overfitting, but also requires more computational resources. By contrast, if wind speed is recognized to affect prices (mainly) because of the resulting wind generation, then using a specialized wind generation forecast (rather than raw wind speed data) appears sensible.[11] A similar reasoning can be applied to other supply and demand components, such as irradiation-driven PV generation, or temperature-driven electric heating.

In summary, three main approaches to incorporate weather data exist. One consists of incorporating a representation of the actual weather state [18]. The price forecasting model can internally capture relationships between the actual and the future weather state. Nevertheless, these relationships are better captured by specialized (ensembled) numerical weather prediction models (NWP). A second approach is therefore to incorporate weather predictions produced by NWP models [17]. A third approach incorporates intermediate forecasts of (weather-driven) quantities that determine the electricity prices, such as load and renewable generation [19]. This last approach embeds the causal relationship between weather and prices explicitly in the model's structure.

Few studies that point out the benefit of intermediate weather-driven quantity forecasts already exist. It is found that producing intermediate wind generation and demand forecasts can lead to improved price forecasts. In the same study [20], it appears that incorporating these forecasts individually leads to better result than using them in combination. Luo and Weng [21] implement a two-stage approach consisting of a first stage that forecasts wind generation and a second one that is trained to map historical wind generation to historical prices. Multiple models are used (polynomial regression, support vector machines, neural networks, deep neural networks) and results indicate that the two-stage approach is consistently superior to the direct approach. However, it is unclear whether similar results can be expected when forecasting prices using a highly dimensional feature space because: (i) the input feature space consists only of historical averaged wind generation and prices, and (ii) the forecasting horizon of one month is connected to a decrease in weather predictability, which is likely to hide the full potential of this approach. Using quantile regression and GARCH, Matsumoto and Endo [22] analyse the effect of introducing an intermediate temperature forecast when predicting week-ahead average prices in Japan. Although these studies highlight the benefits of intermediate forecasts, none uses a feature space that is comprehensive enough to be representative of real-world state-of-the-art day-ahead price forecasts.

---

[10] Relationships clearly remain limited to the modelling approach. A linear model is only able to capture linear relationships.
[11] Replacing wind speed forecasts with wind generation reduces the dimensionality because for the generation it is sufficient to know the aggregate on a bidding zone level, whereas wind speed (being a local phenomenon) needs to be considered with higher geographical granularity.

When demand and renewable generation are forecasted to form the residual load, the only process that is left out is the supply curve which maps the residual load to prices. Some studies focus on the representation of the demand and supply curves in order to forecast prices in Italy and Germany [23], [24]. One key element is, however, that these studies represent demand and supply traded at the exchange, which, in the European context is only a fraction of the system's demand and supply. Theoretically, modelling the quantities at the exchange or total system quantities should lead to the same price prediction because of the possibility to arbitrage. However, quantities at the exchange might be subject to factors that go beyond "structural" drivers, such as the choice to trade on the intraday rather than the day-ahead market. From this perspective, modelling system-wide supply and demand curves appears advantageous.

Furthermore, decomposing the price formation process allows for implementing non-linear transfer functions, such as between wind speed and wind generation, or between residual load and power prices [25]. This mitigates the drawbacks of linear forecasters in the context of high frequency data that are typically more difficult to predict [9]. Encoding domain knowledge into the model's structure by splitting the price forecasting problem into sub-problems facilitates traceability. Moreover, the model used to produce the intermediate forecasts can be customized with regard to the input selection and the modelling approach. However, enforcing this domain knowledge also reduces the model's flexibility to represent "unknown" relationships. This trade-off will be explored and quantified as part of this manuscript.

**Contribution**

In this study I examine the benefits of structuring a statistical electricity price forecasting model. The structure is introduced by:

- replacing raw weather forecasts with individual demand and renewable generation forecasts,
- modelling the market clearing by estimating the supply curve,
- pre-processing wind speed to account for its non-linear relationship with the wind generation.

By performing a ceteris-paribus comparison between a structured model and its unstructured counterpart, this article contributes to the research in the field of wholesale price forecasting by:

- exposing the benefits of domain-knowledge in the context of limited training data yet feature rich models,
- providing quantitative evidence to the more general discussion on the relationship between the forecasting horizon and the modelling paradigm (statistical vs. structural),
- proposing a transparent structured statistical model based exclusively on openly accessible data,
- examining the trade-off between the information lost when using intermediate quantity forecasts, and the estimation uncertainty when highly dimensional weather data are used.

Finally, this study picks up on major trends in the field of day-ahead price forecasting, namely (i) the exploration of non-linear effects (with reasonable computational costs), the use of forecast combinations, and the focus on the supply and demand curves underlying the price formation [26].

## 2  Methods

Three forecasting models are proposed to evaluate the benefit of gradually encoding domain knowledge (see Figure 1). They all used directly or indirectly (i.e. via intermediate quantity forecasts) the same weather predictions to allow for a ceteris paribus comparison. They are modelled as a combination of LASSO Estimated Autoregressive models (LEAR), or as a combination of supply functions. Finally, a fourth model is introduced, to evaluate the benefits of a forecast combination.

  i. **Direct** – Weather forecast attributes including irradiation, cloud cover, wind speed and temperature are paired with weekday and holiday dummies as well as autoregressive components and used directly to forecast the electricity price.
     (For a detailed description see the model variant name *Linear* proposed in [17].)
 ii. **Semi-structured** – Weather forecasts are used to produce intermediate forecasts, namely PV, onshore and offshore wind generation, as well as load and net imports. These intermediate forecasts are paired with weekday and holiday dummies as well as autoregressive components and used to forecast electricity prices.
iii. **Structured** – Summarizes the intermediate forecasts to form a residual load forecast and estimates the supply curve (without renewable generation) based on past observations. The residual load is then transformed using the estimated supply curve to generate the price forecast.
 iv. **Combined** – Combines the aforementioned three models by using a weighted average. The weights are trained based on the past performance.

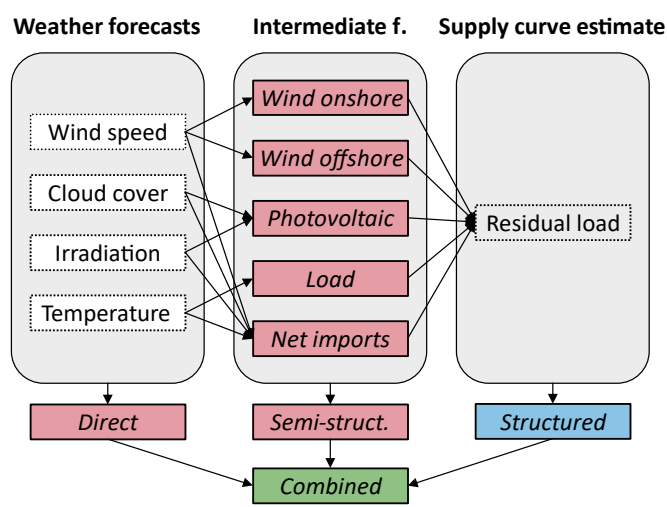

*Figure 1: Model setup comprising a set of LEAR combination (red), an isotonic estimates combination (blue), and a price forecast combination (green).*

The Semi-structured model is similar to studies that use expert models, and incorporate publicly available quantity forecasts in addition to autoregressive and seasonal components [27], [28]. Therefore, seasonal components are kept in this model even if they are already accounted for in the underlying quantity forecasts. On the other hand, the supply curve is estimated as a function that captures the relationship between the price and the residual load. Because the function's domain and codomain are one-dimensional, seasonal components are not incorporated explicitly but only implicitly as part of the residual load (i.e. the underlying constituent quantities). Finally, rather than using external sources, intermediate quantity forecasts are produced as part of this study not only to allow for a ceteris paribus comparison, but also (i) because publicly available quantity forecasts are found to be biased [19] and (ii) to accommodate the extended horizon of 3 days analysed in this study.

The following sections introduce the input data used (Section 2.1), the data selected as input for the intermediate quantity forecasts (Section 2.2), the adopted modelling approach and the respective formulation (Section 2.3), and clarifications on the setup of the experiment (Section 2.4).

## 2.1 Data sources

For the sake of reproducibility, only freely available data are used. These data include weather forecasts from the German weather institute DWD, as well as day-ahead prices and system-wide quantities from ENTSO-E Transparency.

Fuel futures and electricity forward prices are not included because, to the best of my knowledge, there is no publicly accessible provider that regularly updates these prices whilst distributing this data without license restrictions. Commodity prices normally fluctuate only moderately within and between days, because unlike electricity, the underlying commodity is easier to store. Therefore, lagged prices can be sufficient to capture the effect of slowly changing commodity prices. This argument is used in earlier literature [15] that does not incorporate such prices. This consideration is reinforced by the results of a more recent experiment [17], where commodity prices are removed by the LEAR selection in the two parameter-rich variants. However, this caveat is relevant since part of the experiment presented in this study had been conducted during the European gas crisis. During this period, abruptly changing expectations caused extreme gas price fluctuations. Under these circumstances, explicitly incorporating gas prices might have further improved the prediction accuracy.

## 2.2 Input pre-processing

For each intermediate quantity forecast, inputs can be pre-selected, since not all quantities need to be forecasted using the full set of inputs. **PV generation** is mostly correlated with solar irradiation [29]. Paired with cloud cover, these two features are collected for the locations in Germany to predict the PV infeed. **Onshore and offshore wind generation** is mainly (although not exclusively) driven by the wind speed. The relationship between wind speed and wind generation is non-linear because of the non-linear power curve that characterizes wind turbines, and because of the non-linear relationship between wind speed and the altitude. Therefore, the input wind speed that corresponds to an altitude of 10m is transformed to a representative rotor height using the logarithmic wind profile law [30] and the roughness length mapping provided by the Copernicus Land Monitoring Service [31]. The resulting onshore and offshore wind speed is further transformed using the power curves of respectively the Vestas V126 turbine and the Aerodyn SCD turbine.[12] This is a common procedure [33], [34], exception made for that in this manuscript location weights are determined as part of the regression, and not a priori determined by the distribution of wind farms.[13] It is also worth mentioning that bypassing the need for weather re-analyses might avoid substantial biases [36].[14] **Electricity demand** is crucially influenced by temperature because electricity is used to provide space heating and cooling, whereas other weather attributes such as wind speed and irradiation appear to play a minor role [38]. Unlike PV and wind generation, demand is also influenced by calendar effects that can be encoded in the form of weekday and holiday dummy variables. Domestic generation and demand are assumed to be driven by weather conditions in Germany, which allows to further narrow down the selection of used weather locations. Finally, using a domestic supply curve makes only sense if **net imports** from neighbouring countries are considered. In reality, net imports are the result of price differentials between interconnected markets. Instead of using forecasted prices (that are obviously not available at this stage), net imports are modelled with the underlying price drivers. Therefore, all weather attributes across Europe as well as calendar dummies are included.

---

[12] Both power curves are available on the site https://wind-turbine-models.com/ (last accessed on: 01.11.2022). The Vestas V126 is the most common onshore turbine in Germany in 2020 [32], whereas the Aerodyn SCD is the only 8 MW offshore power curve available on the webpage.

[13] An interesting point discussed in [35] is that the transformation neglects the error component of the meteorological variable, which is strictly speaking incorrect "even though it may still be appropriate for point prediction purposes where error measures are calculated against noisy response data".

[14] Instead of using representative wind-power curves, it would also be possible to estimate the "systems" wind-power curve, plenty of approaches exist [37]. However, this would go beyond the scope of this paper.

## 2.3 Models

The study uses three types of regressions: LEAR for feature selection (Section 2.3.1), non-negative least squares regression for model combination with fixed sum coefficients (ensembler, Section 2.3.2), and weighted isotonic regression for an estimation of the supply curve (Section 2.3.3). These components are justified and summarized below.

### 2.3.1 LEAR

In a comprehensive comparison [9]**,** the autoregressive model with exogenous variables called "fARX" results to be the most accurate statistical model, and even better than some machine learning models. This is imputed to the automatic feature selection and the ability to process a large number of inputs. This model, which had been proposed already in a previous study [27], consists of a linear least-squares regression, complemented by a regularization penalty. As pointed out in a review of state-of-the-art algorithms [3], the LASSO estimated autoregressive model (LEAR), which is equivalent to the aforementioned "fARX", emerges among the statistical approaches as the benchmark to beat. Also deep neural networks are positioned as benchmarks to beat; however they are data intensive. In fact, although both models use 250 features, the LEAR is tested on training samples between 8 weeks and 4 years. On the other hand, a training set shorter than 4 years is not even considered in the case of the deep neural network.[15]

Another factor that favours LEAR over other machine learning methods is speed. The ability to re-train the model when new data becomes available has the obvious advantage that it allows for the model to adapt to changing market conditions. However, retraining the model poses severe constraints on the acceptable computation time – an aspect which appears to be left aside by the literature on day-ahead price forecasting. For example, a weather forecast published in the morning can be used to produce a price forecast if all necessary calculations are completed before the day-ahead auction closes, i.e. within few hours**.** This favours LEAR also over other feature selection methods required to avoid overfitting.[16] In fact, embedded methods such as regularization are computationally less expensive than wrapped methods, whilst still being able to consider the model during the feature selection process [2].[17] Examples of machine learning models that are re-train daily exist, such as a DNN that is re-trained daily using a 4 year rolling window [3]. However, this model uses "only" 250 features and there is no indication about the time required by the recalibration.

The following equations summarize the LEAR model.

$$y_{t,h} = \mathbf{x}_t^T \boldsymbol{\beta}_{t,h} + \varepsilon_{t,h}^{(L)} \tag{1}$$

$$\widehat{\boldsymbol{\beta}}_{t,h}^{(\lambda)} = \arg\min_{\boldsymbol{\beta}} \frac{1}{M^{(L)}} \left\| \mathbf{X}_t \boldsymbol{\beta} - \mathbf{y}_{t,h} \right\|_2^2 + \lambda \|\boldsymbol{\beta}\|_1 \tag{2}$$

---

[15] It is worth noting that models are understood as feature-rich when already dozens/hundreds of features are used
[16] Methods to perform feature selection can be categorized in filter, wrapper, and embedded methods [39].
[17] This capability is not preserved by filter methods that do not consider the specific model and, thus, the modelled interaction between features.

$$\hat{y}_{t,h}^{(\lambda)} = \boldsymbol{x}_t^{\mathrm{T}} \hat{\boldsymbol{\beta}}_{t,h}^{(\lambda)} \tag{3}$$

For a given regularization penalty $\lambda$, the LEAR determines the coefficients that correspond to the LASSO complemented least squared error of $M^{(L)} = 84$ observations, $\boldsymbol{X}_t \coloneqq [\boldsymbol{x}_{t-1} \quad \cdots \quad \boldsymbol{x}_{t-M^{(L)}}]^{\mathrm{T}}$, which corresponds to 12 weeks of data. The prediction $\hat{y}_{t,h}^{(\lambda)}$ is a function of the regularization penalty $\lambda$. This is a simplified description that omits the standardization of features as well as the dependency of the training data $\boldsymbol{X}_t$ on the forecasting horizon (for a detail description consult [17]).

### 2.3.2 Ensembler

Model combinations and their benefits have been part of the scientific literature on electricity system forecasting for decades [40]. When models are combined linearly, the corresponding weights can be set all equal, reflecting a naïve approach, or chosen based on the past performance. In the latter case, it is possible to constrain the sum of wights to equal 1. As highlighted in [41], the unconstrained combination comes at the cost of a slower convergence, which closely relates to the estimation uncertainty associated with the higher flexibility. Interestingly, it is beneficial to combine models even if they only differ in the calibration window [4], [42], [43] – combination which leads to better results than selecting ex-post the optimal calibration window.[18] The reason behind ensemble forecasts exhibiting an improved performance is that the results at the outer percentiles tend to be more stable [44]. Analogously, it seems reasonable to expect that combining LEAR models with different regularization penalties might be more accurate that estimating the best penalty ex-ante.

The following expressions describe the ensembler, i.e. the model that combines forecasts based on their past performance.

$$y_{t,h} = \hat{\boldsymbol{y}}_{t,h}^{\mathrm{T}} \boldsymbol{\omega}_{t,h} + \varepsilon_{t,h}^{(E)} \tag{4}$$

$$\hat{\boldsymbol{\omega}}_{t,h} = \arg\min_{\boldsymbol{\omega}} \| \hat{\boldsymbol{Y}}_{t,h} \boldsymbol{\omega} + b - \boldsymbol{y}_{t,h} \|_2^2 \tag{5}$$

$$\hat{y}_{t,h} = \hat{\boldsymbol{y}}_{t,h}^{\mathrm{T}} \hat{\boldsymbol{\omega}}_{t,h} \tag{6}$$

s.t.

$$\sum_{i \in \{1,\ldots,N\}} \hat{\omega}_{t,h}^{(i)} = 1 \tag{7}$$

$$\hat{\omega}_{t,h}^{(i)} \geq 0, \qquad \forall\, i \in \{1, \ldots, N\} \tag{8}$$

The vector $\hat{\boldsymbol{y}}_{t,h} \coloneqq \left[\hat{y}_{t,h}^{(\lambda_1)} \quad \cdots \quad \hat{y}_{t,h}^{(\lambda_N)}\right]^{\mathrm{T}}$ contains the estimations of individual LEAR models that differ only by the regularization penalty $\lambda$. In this study we used $N = 5$ different penalties, ranging from a value that causes overfitting to one that causes underfitting.[19] The individual LEAR forecasts are correlated. However, weights are constrained to positive values and to sum-up to 1 to deal with multicollinearity and reflect the idea of a stacking model. The estimated weights correspond to the least squared error of $M^{(E)} = 100$ past observations $\hat{\boldsymbol{Y}}_{t,h} \coloneqq [\hat{\boldsymbol{y}}_{t-1,h} \quad \cdots \quad \hat{\boldsymbol{y}}_{t-M^{(E)},h}]^{\mathrm{T}}$, i.e. 20 observations per degree of freedom ($N$).

### 2.3.3 Isotonic regression

To model the mapping between the forecasted residual load and price estimates, a functional factor model consisting of a set of time invariant base functions weighted by a time variant coefficient has been proposed in a previous study using 2006-2008 data [16]. Time invariant base functions assume a certain homogeneity in the function that maps the residual demand to prices. This might have been the case in the past, but this far less true in the period considered in this study that includes the European gas crisis. When considering the merit order model, or for that matter supply curves in general, it becomes apparent that these curves mapping residual load to prices are monotonous. This study will therefore use a simpler approach consisting of a isotonic regression [45]. The following expressions describe this type of non-parametric regression.

---

[18] The interpretation is that models trained on long time slices are better at capturing overall relationships, whereas shorter models are quicker at adapting to structural changes.
[19] The regularization penalties are: $\lambda_1 = 10^{-2}$, $\lambda_2 = 10^{-1.5}$, $\lambda_2 = 10^{-1.0}$, $\lambda_4 = 10^{-0.5}$, and $\lambda_5 = 10^{0.0}$.

$$\hat{f}_t = \arg\min_f \sum_{i=t-1}^{t-M^{(I)}} k_i \left(f(x_i) - y_i\right)^2 \tag{9}$$

s.t.

$$\hat{f}(x_i) \geq \hat{f}(x_j), \quad \forall\, (i,j) : x_i \geq x_j \tag{10}$$

The key restriction is that the estimated function $\hat{f}_t$ must be monotonous (non-decreasing), which is $\hat{f}(x_i) \geq \hat{f}(x_j)$ whenever $x_i \geq x_j$. This regression is used to estimate the supply curve (without renewable generation). More recent observations are more valuable for the estimation due to fluctuating commodity prices, and generation unit availabilities. This can be accounted for using weights $k_i$ that get smaller when training observation are further in the past. Four variations of this regression are included where $k_i$ decays linearly to zero after respectively 1, 2, 4, and 8 weeks. A quicker decay results in a supply curve that adjusts more promptly to changing market conditions, whereas a slower decay allows for more observations, thus a more stable regression. These models are combined, i.e. stacked, analogously to the LEAR models.

## 2.4 Experiment overview

As part of this study, several forecasts are produced and compared. Weather forecasts are used as input to produce a price forecast directly (Direct) as well as intermediate quantity forecasts. Each of these forecasts is produced with a set of 5 multivariate LEAR models (differing by the regularization penalty, Section 2.3.1) that are combined with a non-negative fixed sum ensembler (Section 2.3.2). The same procedure is carried out when intermediate forecasts are used to produce a quantity-based price forecast (Semi-structured). The intermediate forecasts are also combined to form a residual load estimate[20] and produce a third forecast (Structured). This forecast consists of four isotonic functions (differing by the length of the training period, Section 2.3.3) that are combined using a non-negative ensembler.[21] Finally, the last forecast is produced based on the three aforementioned ones (Combined) using, again, a non-negative fixed sum ensembler. The individual models and their relationship are summarized in Figure 1. On a daily basis, from the 1st of October 2022 to the 30th of April 2023 (211 days), each model is re-trained with the information available on that day and used to produce and hourly forecasting for the next 3 days. It is worth highlighting, that the level of model nesting strongly impacts the need for historical data, since a model may refer to the past performance of another model. To produce a Combined forecast on the 1st of October 2022, mid-2021 data needs to be available.

The models produced in this study are impacted by the European gas crisis. Model training and prediction periods range from the extreme market conditions observed in the second half of 2022 to the more moderate state mid-2023 (see Figure 5). This effect is discussed in detail in Section 3.2.

---

[20] The residual load estimate is simply the load forecasts minus the variable renewable and net import forecasts.
[21] The constraint forcing the coefficients to sum up to 1 is relaxed because the isotonic regression produced a univariate model that might be biased when a particular hour of the day is considered.

## 3  Results

Results are presented in two sections. A first section that highlights the overall accuracy of the presented price models (Section 3.1), and a second section that highlights the impact of varying market conditions on the individual models (Section 3.2). A third section will discuss limitations (Section 3.3).

### 3.1  Forecasts accuracy

Forecasts are evaluated for each hour of the forecasting horizon ($h$) individually based on the RMSE and NRMSE ($N = 221$). The normalization is used to control for the variability that changes during the day, but also to align the magnitude of errors when different quantity forecasts are compared.

$$\text{RMSE}_h = \sqrt{\frac{1}{N} \sum_{t=1}^{N} (\hat{y}_{t,h} - y_{t,h})^2} \qquad (11)$$

$$\text{NRMSE}_h = \frac{\text{RMSE}_h}{\sigma_h}, \quad \sigma_h = \sqrt{\frac{1}{N-1} \sum_{t=1}^{N} (y_{t,h} - \bar{y}_h)} \qquad (12)$$

For all price models, the accuracy varies substantially as a function of the forecasting horizon. The RMSE exhibits a pronounced daily pattern that is driven by the higher price variance during daytime hours. This pattern fades out, or even reverses, when NRMSE is considered. Especially in the first few hours the Structured model notably underperforms compared to the models that contain lagged prices and can thus represent autoregressive effects; simply using a Direct forecast appears to be a competitive approach (see Figure 2). However, after the 6[th] hour the (N)RMSE exposes a substantial improvement when structuring the model to forecast daytime hours. Interestingly, combining the forecasts is rarely better than the best individual price forecast. This is explained by the fact that all price forecasts use the same input data, and an error in the underlying weather prediction is likely to propagate similarly in each price model (error correlation). Also error autocorrelation plays a role (Section 3.2). It is worth noting that previous research [20] found that using wind generation and demand forecasts separately, rather than their aggregation, led to better results. This finding does not generalize to this experiment, since the Structured is partially better than the Semi-structured variant.

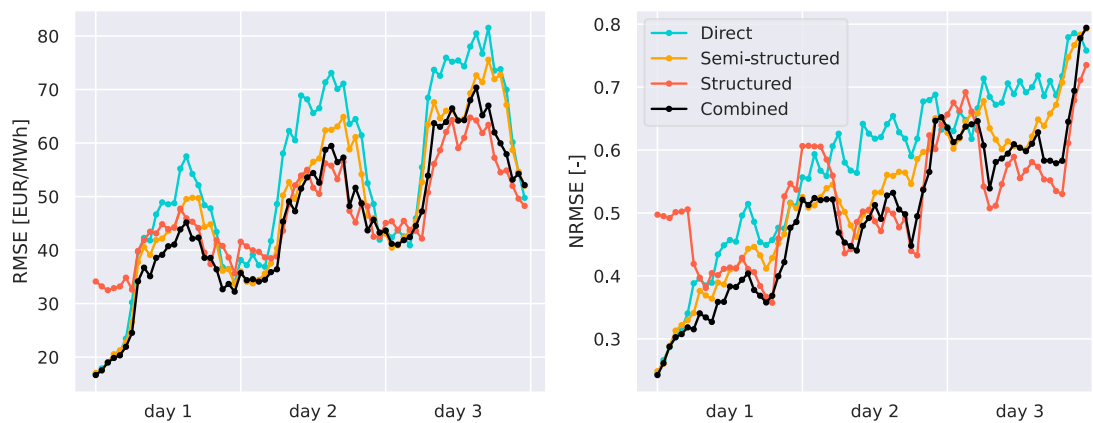

*Figure 2: The forecasting errors (RMSE and NRMSE) of all four price models increase with longer forecasting horizons. Models that leverage autoregressive effects are particularly accurate in the first few hours of the forecasting horizon. Especially daytime prices improve when some kind of structure I imposed.*

Similarly to the accuracy of the price forecasts, quantity forecasts become less accurate at longer forecasting horizons. Unsurprisingly, the residual load is the least accurate in absolute terms (RMSE). However, the relative error (NRMSE) is highest for net imports and offshore wind generation. This reflects the challenge of capturing respectively complex cross-border effects and predicting small quantities. The latter also applies to the solar generation in the early morning and late evening (see Figure 3)

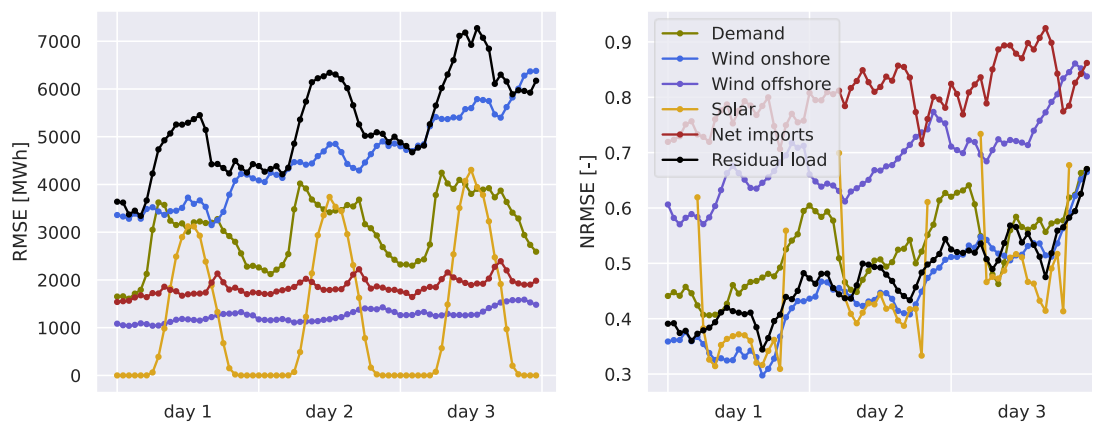

*Figure 3: The quantity forecasts' RMSE exhibit a pronounced daily patter due to the higher volatility during daytime hours. Wind onshore generation is the biggest source of uncertainty (RMSE) although the forecast is comparatively good (NRMSE).*

Finally, the Diebold-Mariano test (DM) – which is commonly used in electricity price forecasting [3] – reveals that the Semi-structured, Structured and Combined forecasts are significantly better than the Direct forecast for daytime prices. The throughout highest DM statistic[22] is achieved by the combined forecast in the first day of the forecasting horizon.

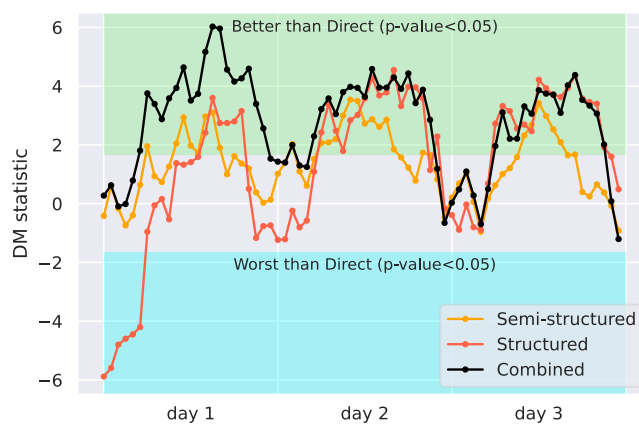

*Figure 4: Diebold-Mariano test shows that the structured model variants provide a statistically significant improvement over the Direct forecast especially when forecasting daytime prices.*

## 3.2   Heterogenous conditions

As a result of the European gas crisis, day-ahead electricity prices spiked in August 2022 and fell to more moderate levels mid-2023. Figure 5 highlights the extent to which day-ahead price fluctuations have been driven by the gas prices, and thus the effect on the supply curve. Although gas prices are not incorporated directly, they are implicitly captured with a certain delay by the lagged prices contained in the Direct and Semi-structured models, and by the estimated supply curves in the Structured variant.

---

[22] High values in the DM statistic are equivalent to low p-values under the hypothesis that the Direct forecast is actually more accurate.

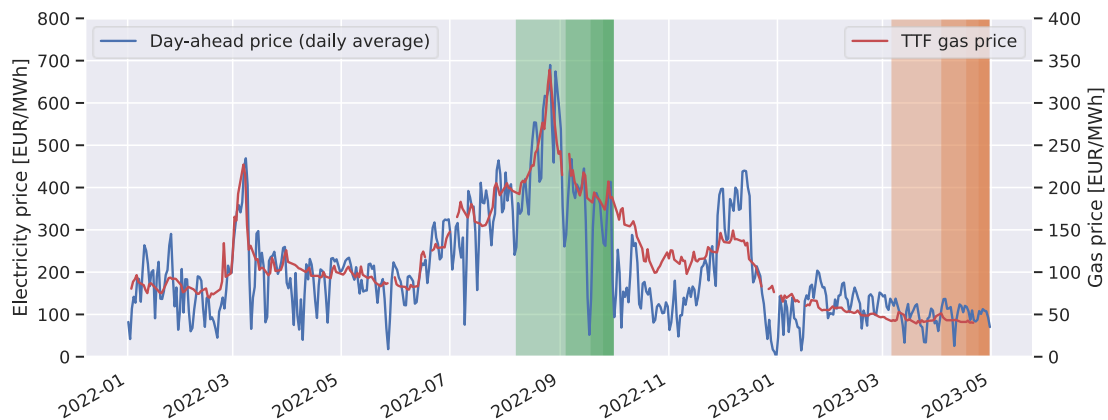

*Figure 5: The training period of the isotonic regressions at the beginning of the experiment (green, the 4 shades highlight the different training lengths) include extreme gas prices that stabilize towards the end of the of the experiment (orange).*

Lagged prices and estimated supply curves do not capture the gas price at the moment when the forecasts are produced, but rather the gas prices preceding this point in time. Shorter training periods increase the estimation uncertainty but contain less outdated gas price levels. This becomes apparent when comparing the training periods of the supply curves (isotonic regression) at the beginning of the experiment (1st October 2022) and at the end of the experiment (30th April 2023). These periods are marked respectively in "green" and "orange" in Figure 5 (the shades representing the four training lengths of the isotonic regressions). In fact, the isotonic ensembler assigns the strongest coefficient to the shortest regression (1 week) on the 1st of October 2022, whereas regressions with longer training periods emerge more strongly on the 30th of April 2023 (see Figure 6).

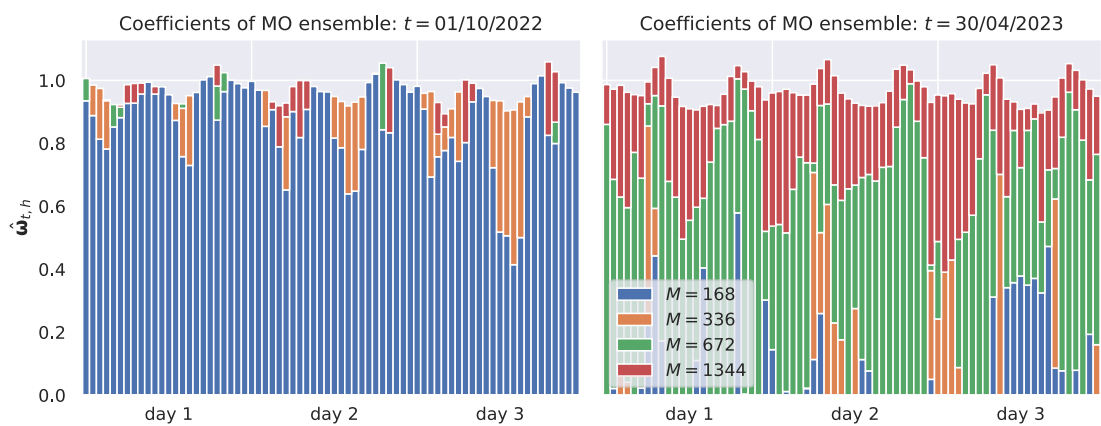

*Figure 6: Supply curves estimated on shorter training periods (1 week) have a high weight at the beginning of experiment because a quickly adapting supply curve estimation is more accurate due to fluctuating market conditions.*

The importance of the training period can also be deduced by the estimated isotonic curves and the corresponding observations shown in Figure 7. Especially the isotonic regression fitted on the 1st of October 2022 on the preceding 8 weeks ($M^{(I)} = 1344$) is substantially different than the other regressions with a shorter training period. On the other hand, the regressions fitted on the 30th of April 2023 are rather homogeneous, reflecting more stable market conditions.

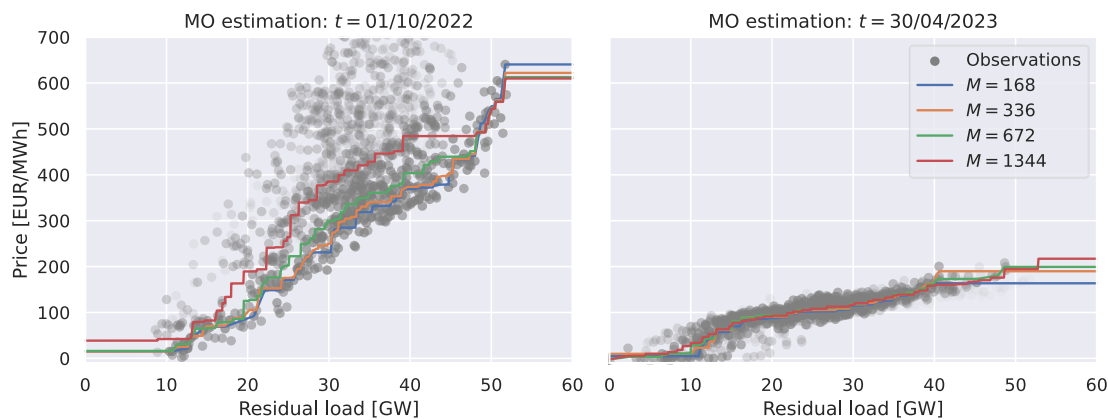

*Figure 7: The estimated supply curves at the beginning and towards the end of the experiment are radically different reflecting the underlying gas price driven market conditions.*

Also the composition of the price models in the Combined model varies over time. During both, the end and beginning of the experiment, models, that incorporate lagged prices exhibit larger weights at the beginning of the forecasting horizon. On the 1st of October 2022 the Combined model is dominated by the Semi-structured variant whereas on the 30th of April 2023 the composition of Direct and fully Structured forecasts plays the bigger role (see Figure 8).

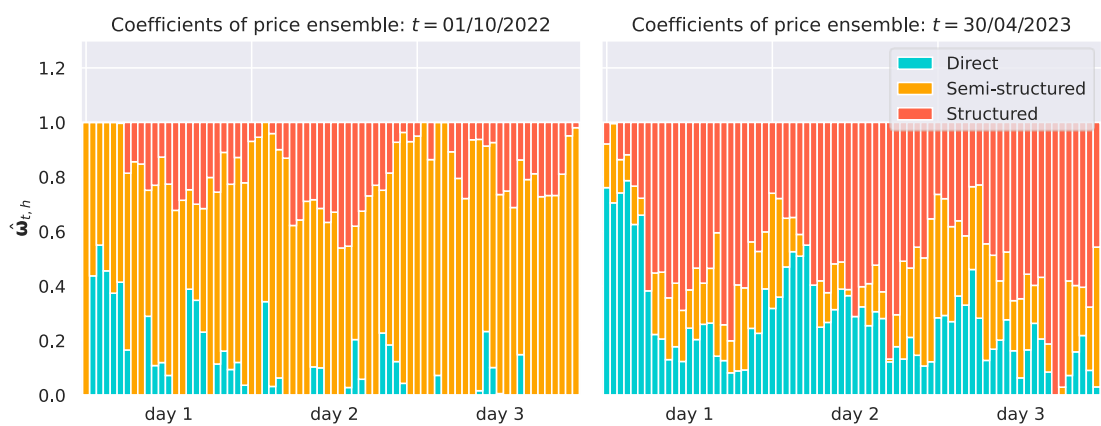

*Figure 8: The weights assigned to models leveraging autoregressive effects (Direct and Semi-structured) are dominant in the first few hours of the forecasting horizon. The Structured model contributes to daytime hours in particular. However the model composition varies greatly depending on the market conditions.*

Finally, it is worth highlighting that the overall accuracy exhibited in Figure 2 does not expose the interesting dynamics that characterized the experiment period. For example, as shown in Figure 9, from mid-November to mid-December the Structured forecast performed substantially worst, probably due to the price spike and subsequent drop registered in the same period. This might indicate a lower resilience to abrupt shocks than the Semi-structured model. These fluctuating conditions lead to error autocorrelation, which can hinder the consistency of the ensembler estimator. Hence, past performance leads, at times, to a suboptimal model combination for forecasting purposes. This finding complements previous studies that conducted experiments in more stable market conditions [4], [42], [44], [46]. As discussed in the previous Section, a further drawback of the Combined forecasts is that weather forecasts are a shared source of errors, thus resulting in error correlation between forecasts.

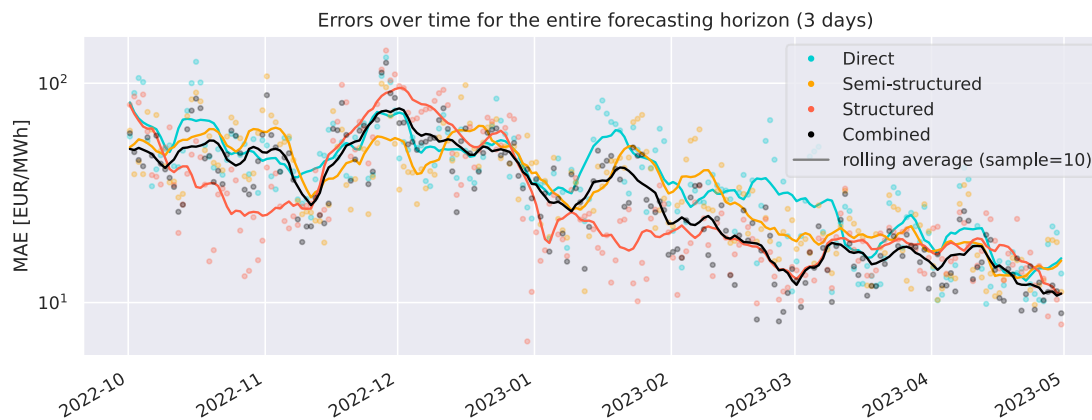

*Figure 9: Depending on the market conditions, absolute errors vary greatly, and no model appears to be throughout better, even not the combined one.*

## 3.3  Limitations

Especially when considering that weather forecasts are a main source for the forecasting error, integrating a probabilistic weather forecast based on NWP ensembles, rather than a single deterministic NWP, would capture part of the uncertainty associated with the forecast [33], and potentially improve the accuracy of derived forecasts. It could also reduce the correlation between forecasts, thus increasing the benefit of using price forecast combinations.

The supply curve is modelled as a function of the price whereas the demand is assumed to be inelastic. This is a simplification motivated by the fact that (i) part of the elasticity is implicitly captured by the load forecast (given that prices correlate with the input features used to forecast the demand), and that (ii) only a moderate – if any – elasticity is attributed to the system-wide electricity demand. However, if the demand elasticity is believed to play a significant role, all features leading to the price formation should be included in the load forecast as well, or a modelling approach that represents the demand as a function of the price should be adopted.

Unlike the estimated supply curve, the LEAR as well as the model ensembles only capture linear relationships between the input feature space and the target variable. The relationship between wind speed and wind generation is known to be non-linear, and this is tackled in this study by pre-transforming the wind speed using representative wind turbine power curves. Other non-linear relationships are not considered.

LEAR ensembles are used to effectively overcome the necessity to choose ex-ante a fixed regularization penalty for the LASSO. This penalty could have been selected adaptively with online methods, such as Bernstein Online Aggregation and Exponentiated Gradient Algorithm [46] or ex-ante with procedures common in the field of hyperparameter tuning, such as Bayesian Optimization [47].

# 4   Conclusions

This study shows that structuring a forecasting model for day-ahead electricity prices can lead to substantial accuracy improvements. These improvements can be attributed to two aspects:

   i. producing quantity forecasts with additional training data that is otherwise not used,
   ii. encoding domain knowledge representing well-understood price formation mechanisms to reduce estimation uncertainty.

This study further provides insights into when structuring forecasts is particularly useful. First, with regard to the forecasting horizon, the Direct forecast is accurate at the beginning, whereas the benefits of structuring the statistical model materialize after the first few hours. Second, adding structure is particularly beneficial to forecast daytime prices. Compared to prices during the night, a larger share of their volatility is captured and explained by intermediate quantity forecast and the estimated supply curves.

The Combined forecast exhibits the best overall DM-statistic when benchmarked against the Direct forecast. However, in contrast to previous studies, this model combination does not generally result in a lower forecasting error. This is because weather forecasts are the main source of uncertainty that is shared across all forecasts. This causes the forecast errors of the Direct, Semi-structured and Structured models to be correlated. Furthermore, market trends result in error autocorrelation; the autocorrelated errors bias the estimation of the weights that are used in the Combined model. Especially the error autocorrelation could be tackled in future studies with more sophisticated combination methods that account for such autoregressive properties.

The comparison with other promising machine learning approaches with a focus on computation time and data requirements is beyond the scope of this study, yet an aspect that deserves further research. The same goes for integrating state-of-the-art probabilistic methods. Adding quantiles of the intermediate quantity forecasts might allow for a better representation of the price formation process.

This study also leverages the advantages of the embedded feature selection performed by the LASSO. Further investigating the suitability and performance of other feature selection methods, including filter and wrapper methods, is particularly relevant in the context of highly dimensional data and dynamic market conditions. The feature space in this study is highly dimensional because of the spatial and temporal distribution of weather. Methods that focus on modelling spatio-temporal relationships are a promising direction for research about electricity price forecasting with weather data. Furthermore, future research could investigate to what extent variance stabilizing transformations of prices [3], [48] and decomposition methods [6] are applicable in the context of structured models, especially when modelling supply and demand curves.

Finally, the decomposition of the price forecast into intermediate quantity forecasts can be interpreted as a sort of "supervised" feature extraction, since target features (i.e. the intermediate quantities) are specified. However, extracting features i.e. splitting the price forecast into intermediate forecasts naturally comes with information loss. This information loss can be mitigated by encoding domain knowledge into the model's loss function, rather than creating intermediate forecasts. As a result, the loss function would include the deviation of intermediate variables (e.g. a combination of wind speed data) from observed ones (e.g. the actual wind generation). This is analogous to adding a custom loss function to an intermediate layer of a neural network. Such a procedure would mitigate the information loss whilst incentivizing the model towards forming known relationships.

# Acknowledgments

Thanks to Lion Hirth and Oliver Ruhnau for the valuable feedback, to Florian Ziel for the insightful discussions, to Clemens Stiewe and Alice Xu for the constructive criticism of the manuscript.